\documentstyle[preprint,pre,aps]{revtex}
\begin{document}
\draft
\title{
Universality of Period Doubling in Coupled Maps
}
\author{Sang-Yoon Kim}
\address{
Department of Physics \\ Kangwon National University\\
Chunchon, Kangwon-Do 200-701, Korea
}
\maketitle
\begin{abstract}
We study the critical behavior of period doubling in
two coupled one-dimensional maps with a single maximum
of order $z$. In particurlar, the effect of the maximum-order $z$ on
the critical behavior associated with coupling is investigated by a
renormalization method.
There exist three fixed maps of the period-doubling renormalization
operator.
For a fixed map associated with the critical behavior at
the zero-coupling critical point, relevant eigenvalues
associated with coupling perturbations
vary depending on the order $z$, whereas they are independent
of $z$ for the other two fixed maps.
The renormalization results for the zero-coupling case are also
confirmed by a direct numerical method.
\end{abstract}
%
% End of Abstract
%
\narrowtext

Universal scaling behavior of period doubling has been found in
one-dimensional (1D) maps with a single maximum of order $z$ $(z>1)$,
\begin{equation}
x_{i+1}=f(x_i)=1-A \; |x_i|^z, \;\;\;z>1.
\label{eq:1DM}
\end{equation}
For all $z >1$, the 1D map (\ref{eq:1DM}) exhibits successive
period-doubling bifurcations as the nonlinearity parameter $A$ is increased.
The period-doubling bifurcation points $A=A_n(z)$ $(n=0,1,2,\dots)$,
at which the $n$th period doubling bifurcation occurs, converge to the
accumulation point $A^*(z)$ on the $A$ axis.
The scaling behavior near the critical point $A^*$ depends on the
maximum-order $z$, i.e., the
parameter and orbital scaling factors, $\delta$ and
$\alpha$, vary depending on $z$
\cite{Feigenbaum,Derrida,Hu,Weele}.
Therefore the order $z$ determines universality classes.

Here we study the critical behavior of period doubling in
a map $T$ consisting of two identical 1D maps coupled symmetrically:
\begin{equation}
   T:\left\{
       \begin{array}{l}
        x_{i+1} = F(x_i,y_i) = f(x_i) + g(x_i,y_i), \\
        y_{i+1} = F(y_i,x_i) = f(y_i) + g(y_i,x_i),
       \end{array}
     \right.
\label{eq:CM}
\end{equation}
where $f(x)$ is a 1D map (\ref{eq:1DM}) with a single maximum
of even order $z$ $(z=2,4,6,\dots)$ at $x=0$, and
$g(x,y)$ is a coupling function. The uncoupled 1D map $f$
satisfies a normalization condition, $f(0) = 1$,
and the coupling function $g$ obeys a condition, $g(x,x) = 0$ for
any $x$.

The quadratic-maximum case $(z=2)$ was previously studied in
Refs.~\cite{Kuznetsov,Kook,Kim1,Kim4,Kim2}.
In this paper, using the renormalization method developed in Ref.~\cite{Kim2},
we extend the results for the $z=2$ case to all even-order cases and
investigate the dependence of the critical behavior
associated with coupling on the order $z$.

The period-doubling renormalization transformation $\cal N$ for a
coupled map $T$ consists of squaring $(T^2)$ and rescaling $(B)$
operators:
\begin{equation}
{\cal N}(T) \equiv B T^2 B^{-1}.
\label{eq:RON}
\end{equation}
Here the rescaling operator $B$ is:
\begin{equation}
B = \left( \begin{array}{cc}
                    \alpha &  \;\;\;    0\\
                    0      &  \;\;\;    \alpha
                   \end{array}
           \right),
\label{eq:SO}
\end{equation}
because we consider only in-phase orbits ($x_i \; = \; y_i$ for all $i$).

Applying the renormalization operator ${\cal N}$ to the coupled map
(\ref{eq:CM}) $n$ times, we obtain the $n$ times renormalized map $T_n$ of
the form,
\begin{equation}
 {T_n}:\left\{
       \begin{array}{l}
        x_{i+1} = {F_n}(x_i,y_i) = {f_n}(x) + {g_n}(x,y), \\
        y_{i+1} = {F_n}(y_i,x_i) = {f_n}(y) + {g_n}(y,x).
       \end{array}
     \right.
\label{eq:RTn}
\end{equation}
Here ${f_n}$ and ${g_n}$ are the uncoupled and coupling parts of the $n$ times
renormalized function $F_n$, respectively. They satisfy the following
recurrence equations \cite{Kim2}:
\begin{eqnarray}
&f_{n+1}(x) = &
 \alpha {f_n}({f_n}({\frac {x} {\alpha}})), \label{eq:RUCFn} \\
&g_{n+1}(x,y) =& {\alpha} {f_n}({f_n}({\frac {x} {\alpha}}) + {g_n}({\frac {x}
{\alpha}},{\frac {y} {\alpha}})) \nonumber \\
&& +{\alpha} {g_n}({f_n}({\frac {x} {\alpha}}) + {g_n}({\frac {x} {\alpha}},
{\frac {y} {\alpha}}),{f_n}({\frac {y} {\alpha}}) \nonumber \\
&&\;\;\;\;\;\;\;\;\;\; +{g_n}({\frac {y} {\alpha}},{\frac {x} {\alpha}}))-
{\alpha} {f_n}({f_n}({\frac {x} {\alpha}})),
\label{eq:RCFn}
\end{eqnarray}
where the rescaling factor $\alpha$ is chosen to preserve the normalization
condition ${f_{n+1}}(0) = 1$, i.e.,
${\alpha} = 1/{f_n}(1)$.
Eqs.\ (\ref{eq:RUCFn}) and (\ref{eq:RCFn}) define a renormalization
operator $\cal R$ of transforming a pair of functions, $(f,g)$;
\begin{equation}
 \left( \begin{array}{c}
                    {f_{n+1}} \\
                    {g_{n+1}}
                   \end{array}
           \right)
= {\cal R}  \left( \begin{array}{c}
                    {f_n} \\
                    {g_n}
                   \end{array}
           \right).
\label{eq:RORn}
\end{equation}

A critical map $T_c$ with the nonlinearity and coupling parameters
set to their critical values is attracted to a fixed map $T^*$ under
iterations of the renormalization transformation $\cal N$,
\begin{equation}
  {T^*}:\left\{
       \begin{array}{l}
        x_{i+1} = {F^*}(x_i,y_i) = {f^*}(x_i) + {g^*}(x_i,y_i), \\
        y_{i+1} = {F^*}(y_i,x_i).
       \end{array}
     \right.
\label{eq:FM}
\end{equation}
Here $({f^*},{g^*})$ is a fixed point of the renormalization operator
$\cal R$ with ${\alpha} = {1/{f^*}(1)}$, which satisfies
$ (f^*,g^*) = {\cal R} (f^*,g^*) $.
Note that the equation for $f^*$ is just the fixed point equation
in the 1D map case. The 1D fixed function $f^*$ varies depending on the order
$z$ \cite{Weele}. Consequently only the equation for the coupling
fixed function $g^*$ is left to be solved.

However it is not easy to directly solve the equation
for the coupling fixed function.
We therefore introduce a tractable recurrence equation for a
``reduced'' coupling function of the coupling function $g(x,y)$ \cite{Kim2},
defined by
\begin{equation}
G(x) \equiv {\displaystyle \left.
{\frac {\partial g(x,y)} {\partial y}}\right|_{y=x}}.
\label{eq:RCFCT}
\end{equation}
Differentiating the recurrence equation (\ref{eq:RCFn}) for $g$ with
respect to $y$ and setting $y=x$, we have
\begin{eqnarray}
G_{n+1}(x) &=& [{f'_n}({f_n}({\frac {x} {\alpha}}))- 2G_{n}({f_n}({\frac {x}
{\alpha}}))] G_{n}({\frac {x} {\alpha}}) \nonumber \\
&& + G_{n}({f_n}({\frac {x} {\alpha}})) {f'_n}({\frac {x} {\alpha}}).
\label{eq:RRE}
\end{eqnarray}
Then Eqs.\  (\ref{eq:RUCFn}) and (\ref{eq:RRE})
define a ``reduced'' renormalization operator $\tilde{\cal R}$
of transforming a pair of functions $(f,G)$:
\begin{equation}
 \left( \begin{array}{c}
                    {f_{n+1}} \\
                    G_{n+1}
                   \end{array}
           \right)
= {\tilde{\cal R}}  \left( \begin{array}{c}
                    {f_n} \\
                    G_{n}
                   \end{array}
           \right).
\label{eq:RRO}
\end{equation}

We look for a fixed point $({f^*},{G^*})$ of ${\tilde {\cal R}}$, which
satisfies $ (f^*,G^*) = {\tilde {\cal R}} (f^*,G^*)$.
Here
$G^*$ is just the reduced coupling fixed function of $g^*$ (i.e.,
${G^*}(x) = {\partial {g^*}(x,y)}/{\partial y}|_{y=x}$).
As in the quadratic-maximum case $(z=2)$ \cite{Kim2}, we find three
solutions for $G^*$:
\begin{eqnarray}
1. \;\;\; {G^*}(x) &=& 0. \label{eq:RFP1} \\
2. \;\;\; {G^*}(x) &=& {1 \over 2} {f^*}'(x). \label{eq:RFP2} \\
3. \;\;\; {G^*}(x) &=& {1 \over 2} [{f^*}'(x) - 1]. \label{eq:RFP3}
\end{eqnarray}
Here the first solution, corresponding to the reduced coupling
fixed function of the zero-coupling fixed function $g^*(x,y)=0$, is
associated with the critical behavior at the zero-coupling critical point,
whereas the second and third solutions dependent on the order $z$ are
associated with the critical behavior at other critical points
\cite{Kim2}.

Consider an infinitesimal reduced coupling perturbation $(0,{\Phi}(x))$ to
a fixed point $({f^*},{G^*})$ of $\tilde {\cal R}$.
We then examine the evolution of a pair of functions, $({f^*}(x),
{G^*}(x)+{\Phi}(x))$ under the reduced renormalization transformation
$\tilde {\cal R}$.
In the linear approximation we obtain a reduced linearized operator
$\tilde {\cal L}$ of transforming a reduced coupling perturbation $\Phi$:
%\widetext
\begin{eqnarray}
{{\Phi}_{n+1}}(x) &=& [{\tilde {\cal L}} {\Phi_n}](x) \nonumber \\
&=& [{f^*}'({f^*}({\frac {x} {\alpha}}))-2{G^*}({f^*}({\frac {x} {\alpha}}))]
{\Phi _n}({\frac {x} {\alpha}}) \nonumber \\
&&+ [{f^*}'({\frac {x} {\alpha}})-2{G^*}({\frac {x} {\alpha}})] {\Phi _n}
({f^*}({\frac {x} {\alpha}})).
\label{eq:LO2}
\end{eqnarray}
Here the prime denotes a derivative.
If a reduced coupling perturbation
${\Phi}^*(x)$ satisfies
\begin{equation}
\nu {{\Phi}^*}(x) = [{\tilde {\cal L}} {\Phi^*}](x),
\label{eq:CEVE}
\end{equation}
then it is called a reduced coupling eigenperturbation with coupling
eigenvalue (CE) $\nu$.

We first show that CEs are independent of the order $z$
for the second and third solutions (\ref{eq:RFP2}) and (\ref{eq:RFP3})
of $G^*(x)$.
In case of the second solution $G^*(x)= {1 \over 2} {f^*}'(x)$,
the reduced linearized operator ${\tilde {\cal L}}$ becomes a null operator,
independently of $z$,
because the right hand side of Eq.\ (\ref{eq:CEVE}) becomes zero.
Therefore there exist no relevant CEs.
For the third case $G^*(x)= {1 \over 2} [{f^*}'(x)-1]$,
Eq.\ (\ref{eq:CEVE}) becomes
\begin{equation}
\nu \Phi^*(x) = {\Phi^*}({\frac {x} {\alpha}}) + {\Phi^*}({f^*}({\frac {x}
{\alpha}})).
\end{equation}
When $\Phi^*(x)$ is a non-zero constant function, i.e.,
$\Phi^*(x) = c$ ($c$~:~non-zero constant),
there exists a relevant CE, $\nu = 2$, independently of $z$.

In the zero-coupling case $G^*(x)=0$, the eigenvalue equation (\ref{eq:CEVE})
becomes
\begin{equation}
\nu \Phi^*(x) = {f^*}'({f^*}({\frac {x} {\alpha}})) {\Phi^*}({\frac {x}
{\alpha}})+{f^*}'({\frac {x} {\alpha}}) {\Phi^*}({f^*}({\frac {x} {\alpha}})).
\label{eq:ZCEVE}
\end{equation}
Relevant CEs of Eq.~(\ref{eq:ZCEVE}) vary depending on the order $z$, as
will be seen below.

An eigenfunction $\Phi^*(x)$ can be separated into two components,
$\Phi^*(x) = \Phi^{*(1)}(x) + \Phi^{*(2)}(x)$ with
$\Phi^{*(1)}(x) \equiv a^*_0 + a^*_1 x + \cdots + a^*_{z-2} x^{z-2}$
and
$\Phi^{*(2)}(x) \equiv a^*_{z-1} x^{z-1} + a^*_z x^z + \cdots~$,
and the 1D fixed function $f^*$ is a polynomial in $x^z$, i.e.,
$f^*(x) = 1 + c^*_z x^z + c^*_{2z} x^{2z} + \cdots~$.
Substituting the functions $\Phi^*$, $f^*$ and ${f^*}'$ into the
eigenvalue equation (\ref{eq:ZCEVE}), it has the structure
\begin{equation}
{\nu} a^*_k = {\sum_l} M_{kl}(\{c^*\}) a^*_l,\;k,l=0,1,2,\dots~.
\label{eq:EVE}
\end{equation}
We note that each $a^*_l$ $(l=0,1,2,\dots)$ in the first and
second terms in the right hand side of Eq.\ (\ref{eq:ZCEVE}) is involved
only in the determination of coefficients of monomials $x^k$ with
$k=l+mz$ and $k=(z-1)+mz$ ($m=0,1,2,\dots$), respectively.
Therefore any $a^*_l$ with $l \geq z-1$ (in the right hand side) cannot
be involved in the determination of coefficients of monomials $x^k$
with $k < z-1$, which implies that the eigenvalue equation
(\ref{eq:EVE}) is of the form
\begin{equation}
\nu
\left( \begin{array}{c}
                    \Phi^{*(1)} \\
                    \Phi^{*(2)}
                   \end{array}
           \right)
= \left( \begin{array}{cc}
           M_1 & \;\;\; 0\\
           M_3 & \;\;\; M_2
           \end{array}
           \right) \;
           \left( \begin{array}{c}
                    \Phi^{*(1)} \\
                    \Phi^{*(2)}
                   \end{array}
              \right),
\label{eq:EVE2}
\end{equation}
where $M_1$ is a $(z-1) \times (z-1)$ matrix,
$\Phi^{*(1)} \equiv (a^*_0,\dots,a^*_{z-2})$, and
$\Phi^{*(2)} \equiv (a^*_{z-1},a^*_z,\dots).$
 From the reducibility of the matrix $M$ into a semi-block form, it follows
that to determine the eigenvalues of $M$ it is
sufficient to solve the eigenvalue problems for the two submatrices
$M_1$ and $M_2$ independently.

We first solve the eigenvalue equation of $M_1$
($\nu \Phi^{*(1)} = {M_1} \Phi^{*(1)}$), i.e.,
\begin{equation}
{\nu} a^*_k = {\sum_l} M_{kl}(\{c^*\}) a^*_l,\;\;k,l=0,\dots,z-2.
\end{equation}
Note that this submatrix $M_1$ is diagonal.
Hence its eigenvalues are just the diagonal elements:
\begin{equation}
{\nu}_k = M_{kk} = {\displaystyle {\frac {{f^*}'(1)} {\alpha^k} } }
={\alpha}^{z-1-k},
\;\;\;k=0,\dots,z-2.
\label{eq:EV1}
\end{equation}
Notice that all $\nu_k$'s are relevant eigenvalues.

Although $\nu_k$ is also an eigenvalue of $M$, $(\Phi^{*(1)}_k,0)$
cannot be an eigenvector of $M$, because there exists a third submatrix
$M_3$ in $M$ [see Eq.\ (\ref{eq:EVE2})].
Therefore an eigenfunction $\Phi^*_k(x)$ in Eq.\ (\ref{eq:ZCEVE})
with eigenvalue $\nu_k$
is a polynomial with a leading monomial of degree $k$, i.e.,
$\Phi^{*}_k(x) = \Phi^{*(1)}_k(x) + \Phi^{*(2)}_k(x)
= a^*_k x^k + a^*_{z-1} x^{z-1}+ a^*_z x^z+ \cdots~$,
where $a^*_k \neq 0$.

We next solve the eigenvalue equation of $M_2$
($\nu \Phi^{*(2)} = {M_2} \Phi^{*(2)}$), i.e.,
\begin{equation}
{\nu} a^*_k = {\sum_l} M_{kl}(\{c^*\}) a^*_l,\;\;k,l=z-1,z,\dots~.
\end{equation}
Unlike the case of $M_1$, $(0,\Phi^{*(2)})$ can be an eigenvector
of $M$ with eigenvalue $\nu$. Then its corresponding
function $\Phi^{*(2)}(x)$ is an
eigenperturbation with eigenvalue $\nu$, which satisfies
Eq.\ (\ref{eq:ZCEVE}).
One can easily see that $\Phi^{*(2)}(x) =  {f^*}'(x)$ is an eigenfunction
with CE $\nu = 2$, which is the $z$th relevant CE in addition to
those in Eq.\ (\ref{eq:EV1}).
It is also found that there exist an infinite number of additional
(coordinate change) eigenfunctions
$\Phi^{*(2)}(x) ={f^*}'(x) [f^{*n}(x)-x^n]$ with irrelevant CEs
$\alpha^{-n}$  $(n=1,2,\dots)$, which are associated with coordinate
changes \cite{Kim3}. We conjecture that together with the $z$
(non-coordinate change) relevant CEs, they give the whole
spectrum of the reduced linearized operator
${\tilde {\cal L}}$ of Eq.\ (\ref{eq:LO2}) and the spectrum is
complete.

Consider an infinitesimal coupling perturbation $g(x,y)$
$[= \varepsilon \varphi(x,y) ]$
to a critical map at the zero-coupling critical point, in which case
the map $T$ of Eq.\ (\ref{eq:CM}) is of the form,
\begin{equation}
   T:\left\{
       \begin{array}{lll}
  x_{i+1} \! &=& \! F(x_i,y_i)
  = f_{A^*}(x_i) + \varepsilon \varphi(x,y), \\
        y_{i+1}\! &=& \! F(y_i,x_i),
       \end{array}
     \right.
\label{eq:CM2}
\end{equation}
where the subscript $A^*$ of $f$ denotes the critical value of the
nonlinearity parameter $A$ and $\varepsilon$ is an infinitesimal coupling
parameter. The map $T$ for $\varepsilon=0$ is just the zero-coupling
critical map consisting of two uncoupled 1D critical maps $f_{A^*}$.
It is attracted to the zero-coupling fixed map (consisting of two 1D fixed
maps $f^*$) under iterations of the renormalization transformation
${\cal N}$ of Eq.\ (\ref{eq:RON}).

The reduced coupling function $G(x)$ of
$g(x,y)$ is given by [see Eq.\ (\ref{eq:RCFCT})]
\begin{equation}
G(x)= \varepsilon \Phi(x) \equiv \varepsilon {\displaystyle \left.
{\frac {\partial \varphi(x,y)} {\partial y}}\right|_{y=x}}.
\label{eq:RCFCT2}
\end{equation}
Then $\varepsilon \Phi(x)$ corresponds to an infinitesimal perturbation
to the reduced zero-coupling fixed function $G^*(x)=0$ of Eq.\ (\ref{eq:RFP1}).
The $n$th image $\Phi_n$ of $\Phi$ under the reduced
linearized operator $\tilde {\cal L}$ of Eq.\ (\ref{eq:LO2}) has the form,
\begin{eqnarray}
\Phi_n (x) &=& [{\tilde{\cal L}}^n \Phi](x) \nonumber \\
&\simeq& {\sum_{k=0}^{z-2}} {\alpha_k} {\nu_k^n} {\Phi^*_k(x)} +
{\alpha_{z-1}} {2^n} {f^*}'(x)\;\;{\rm for}\;\;{\rm large}\;\;n, \nonumber \\
&&
\end{eqnarray}
since the irrelevant part of $\Phi_n$ becomes negligibly small for
large $n$.

The stability multipliers $\lambda_{1,n}$ and $\lambda_{2,n}$ of
the $2^n$-periodic orbit of the map $T$ of Eq.\ (\ref{eq:CM2}) are
the same as those of the fixed point of the $n$ times renormalized map
${\cal N}^n (T)$ \cite{Kim2}, which are given by
\begin{equation}
\lambda_{1,n} = {f_n}'({\hat x}_n), \;\;\;
\lambda_{2,n} = {f_n}'({\hat x}_n)- 2 {G_n}({\hat x}_n).
\label{eq:MULTI2}
\end{equation}
Here $(f_n,G_n)$ is the $n$th image of $(f_{A^*},G)$ under the reduced
renormalization transformation $\tilde{\cal R}$
[i.e., $(f_n,G_n) = {\tilde {\cal R}}^n (f_{A^*},G)$], and
${\hat x}_n$ is just the fixed point of $f_n(x)$ [i.e.,
$ {\hat x}_n={f_n}({\hat x}_n)$] and converges to the fixed point $\hat x$
of the 1D fixed map $f^*(x)$ as $n \rightarrow \infty$.
In the critical case ($\varepsilon = 0$), $\lambda_{2,n}$ is equal to
$\lambda_{1,n}$ and they converge to the 1D critical stability
multiplier $\lambda^* = {f^*}'(\hat x)$.
Since $G_n(x) \simeq  [{\tilde {\cal L}}_2^n G](x) = \varepsilon \Phi_n(x)$ for
infinitesimally small $\varepsilon$, $\lambda_{2,n}$ has the form
\begin{eqnarray}
{\lambda_{2,n}} & \simeq & {\lambda_{1,n}} -2 \varepsilon \Phi_n \nonumber \\
 &\simeq& {\lambda^*} + {\varepsilon}
\left[ {\sum_{k=0}^{z-2}} e_k {\nu_k^n} + e_{z-1} {2^n} \right]
\;\;{\rm for\;\;large}\;\;n,
\end{eqnarray}
where ${e_k}=-2{\alpha_k} {\Phi^*_k({\hat x})}\;\;(k=0,\dots,z-2)$
and ${e_{z-1}}=-2{\alpha_{z-1}} {f^*}'({\hat x})$.
Therefore the slope $S_n$ of $\lambda_{2,n}$ at the zero-coupling point
($\varepsilon=0$) is
\begin{equation}
S_n \equiv
\left. {\displaystyle {\frac {\partial \lambda_{2,n}}
{\partial \varepsilon} } }\right|_{\varepsilon=0}
\label{eq:slope}
\simeq {\sum_{k=0}^{z-2}} e_k {\nu_k^n} + e_{z-1} {2^n}\;\;{\rm for\;\;large}
\;\;n.
\end{equation}
Here the coefficients $\{ e_k\;;\;k=0,\dots,z-1 \}$ depends on the
initial reduced function $\Phi(x)$, because $\alpha_k$'s are determined only
by $\Phi(x)$. Note that the magnitude of slope $S_n$ increases with $n$
unless all $e_k$'s ($k=0,\dots,z-1)$ are zero.

We choose monomials $x^l$ $(l=0,1,2,\dots)$ as initial reduced functions
$\Phi(x)$, because any smooth
function $\Phi(x)$ can be represented as a linear combination
of monomials by a Taylor series.
Expressing $\Phi(x) = x^l$ as a linear combination
of eigenfunctions of ${\tilde {\cal L}}_2$, we have
\begin{eqnarray}
\Phi(x) = x^l
&=& {\alpha_l} {\Phi^*_l(x)} +{\alpha_{z-1}}
{f^*}'(x) \nonumber \\
&& +{\sum_{n=1}^{\infty}} {\beta_n} {f^*}'(x) [f^{*n}(x)-x^n],
\end{eqnarray}
where $\alpha_l$ is non-zero for $l<z-1$ and zero for $l \geq z-1$, and
all $\beta_n$'s are irrelevant components.
Therefore the slope $S_n$ for large $n$ becomes
\begin{equation}
S_n \simeq \left\{ \begin{array}{l}
           e_l {\nu_l^n} + e_{z-1} 2^n\;\;{\rm for}\;\; l<z-1, \\
                       e_{z-1} 2^n\;\;{\rm for}\;\; l \geq z-1.
                  \end{array}
           \right.
\label{eq:slope2}
\end{equation}
Note that the growth of $S_n$ for large $n$ is governed by two CEs $\nu_l$
and $2$ for $l<z-1$ and by one CE $\nu=2$ for $l \geq z-1$.

We numerically study the quartic-maximum case $(z=4)$ in the
two coupled 1D maps (\ref{eq:CM2}) and confirm the renormalization
results (\ref{eq:slope2}).
In this case we follow the periodic orbits of period $2^n$ up to
level $n=15$ and obtain the slopes $S_n$ of Eq.\ (\ref{eq:slope}) at the
zero-coupling critical point $(A^*,0)$ $(A^*=1.594\,901\,356\,228\dots)$
when the reduced function $\Phi(x)$ is a monomial $x^l$ ($l=0,1,\dots$).

The renormalization result implies that the
slopes $S_n$ for $l \geq z-1$ obey a one-term scaling law asymptotically:
\begin{equation}
S_n = d_1 r_1^n.
\label{eq:OTSL}
\end{equation}
We therefore define the growth rate of the slopes as follows:
\begin{equation}
r_{1,n} \equiv {\frac {S_{n+1}} {S_n} }.
\end{equation}
Then it will converge to a constant $r_1$ as $n \rightarrow \infty$.
A sequence of $r_{1,n}$ for $\Phi(x)=x^3$ is shown in
the second column of Table I. Note that it converges fast
to $r_1 = 2$. We have also studied several other reduced-coupling cases
with $\Phi(x)=x^l$ $(l>3)$. In all higher-order cases
studied, the sequences of $r_{1,n}$ also converge fast to $r_1 =2$.

When $l<z-1$, two relevant CEs govern the growth of the slopes $S_n$.
We therefore extend the simple one-term scaling law (\ref{eq:OTSL}) to a
two-term scaling law:
\begin{equation}
S_n = d_1 r_1^n + d_2 r_2^n,\;\;{\rm for\;\;large}\;\;n,
\label{eq:TTSL}
\end{equation}
where $ \left| r_1 \right| > \left| r_2 \right|.$
This is a kind of multiple-scaling law \cite{Mao}.
Eq.\ (\ref{eq:TTSL}) gives
\begin{equation}
S_{n+2} = t_1 S_{n+1}  - t_2 S_n,
\label{eq:TTRE}
\end{equation}
where $t_1 = r_1 + r_2$ and $t_2 = r_1 r_2$.
Then $r_1$ and $r_2$ are solutions of the following quadratic
equation,
\begin{equation}
r^2 - t_1 r + t_2 = 0.
\label{eq:EVEr}
\end{equation}
To evaluate $r_1$ and $r_2$, we first obtain $t_1$ and $t_2$
from $S_n$'s using Eq.\ (\ref{eq:TTRE}):
\begin{equation}
t_1 = {\frac { S_{n+1} S_n - S_{n+2} S_{n-1} } {S_n^2-S_{n+1} S_{n-1}} },\;\;
t_2 = {\frac { S_{n+1}^2 - S_n S_{n+2} } {S_n^2-S_{n+1} S_{n-1}} }.
\label{eq:t12}
\end{equation}
Note that Eqs.\ (\ref{eq:TTSL})-(\ref{eq:t12}) are valid for large $n$.
In fact, the values of $t_i$'s and $r_i$'s $(i=1,2)$ depend on the level $n$.
Thus we denote the values of $t_i$'s in Eq.\ (\ref{eq:t12})
explicitly by $t_{i,n-1}$'s, and the values of $r_i$'s obtained
from Eq.\ (\ref{eq:EVEr}) are also denoted by $r_{i,n-1}$'s.
Then each of them converges to a constant as $n \rightarrow \infty$:
\begin{equation}
\lim_{n \rightarrow \infty} t_{i,n} = t_i,\;\;
\lim_{n \rightarrow \infty} r_{i,n} = r_i,\;\;i=1,2.
\end{equation}

The two-term scaling law (\ref{eq:TTSL}) is very well-obeyed.
Sequences $r_{1,n}$ and $r_{2,n}$ for $\Phi(x)=1$ are shown in the third and
fourth columns of Table \ref{table1}.
They converge fast to $r_1 = {\alpha}^3$ $(\alpha=-1.6903 \dots)$
and $r_{2,n}=2$, respectively. We have also studied two other reduced-coupling
cases with $\Phi(x)=x^l$ $(l=1,2)$. It is found that the sequences
$r_{1,n}$ and $r_{2,n}$ for $l=1(2)$ converge fast to their limit values
$r_1 = {\alpha}^2(2)$ and $r_2=2(\alpha),$ respectively.

\acknowledgments
This work was supported by Non-Directed Research Fund, Korea Research
Foundation, 1993.

%
%  End of Reference
%
\begin{table}
\caption{In the quartic-maximum case $(z=4)$, a sequence $\{ r_{1,n} \}$ for a
one-term scaling law is shown in the second column when $\Phi(x)=x^3$, and two
sequences $\{ r_{1,n} \}$ and $\{ r_{2,n} \}$ for a two-term scaling law are
shown in the third and fourth columns when $\Phi(x)=1$.}
\begin{tabular}{cccc}
& \multicolumn{1}{c}{$\Phi(x)=x^3$} & \multicolumn{2}{c}{$\Phi(x)=1$} \\
$n$  & $r_{1,n}$ & $r_{1,n}$ &  $r_{2,n}$ \\
\tableline
5 & 1.999\,920\,2 & -4.829\,455\,8 & 1.958 \\
6 & 2.000\,009\,3 & -4.829\,422\,6 & 2.090 \\
7 & 1.999\,994\,7 & -4.829\,409\,8 & 1.973 \\
8 & 2.000\,000\,4 & -4.829\,406\,8 & 2.039 \\
9 & 1.999\,999\,6 & -4.829\,405\,8 & 1.984 \\
10 & 2.000\,000\,0 & -4.829\,405\,5 & 2.018 \\
11 & 2.000\,000\,0 & -4.829\,405\,5 & 1.992 \\
12 & 2.000\,000\,0 & -4.829\,405\,4 & 2.009 \\
\end{tabular}
\label{table1}
\end{table}
%
%  End of Tables
%

\begin{references}
\bibitem{Feigenbaum} M.\ J.\ Feigenbaum, J.\ Stat.\ Phys.\ {\bf 19}, 25
 (1978); {\bf 21}, 669 (1979).
\bibitem{Derrida} B.\ Derrida, A.\ Gervois, and Y.\ Pomeau, J.\ Phys.\ A
 {\bf 12}, 269 (1979).
\bibitem{Hu} B.\ Hu and I.\ I.\ Satija, Phys.\ Lett.\ A {\bf 98}, 143 (1982).
\bibitem{Weele} J.\ P.\ van der Weele, H.\ W.\ Capel, and R.\ Kluiving,
 Physica A {\bf 145}, 425 (1987).
\bibitem{Kuznetsov} S.\ Kuznetsov, Radiophys. Quantum Electron.
{\bf 28}, 681 (1985).
\bibitem{Kook} H.\ Kook, F.\ H.\ Ling, and G.\ Schmidt, Phys.\ Rev.\ A
{\bf 43}, 2700 (1991).
\bibitem{Kim1} S.-Y.\ Kim and H.\ Kook, Phys.\ Rev.\ A {\bf 46}, R4467 (1992).
\bibitem{Kim4} S.-Y.\ Kim and H.\ Kook, Phys.\ Lett.\ A {\bf 178}, 258 (1993).
\bibitem{Kim2} S.-Y.\ Kim and H.\ Kook, Phys.\ Rev.\ E {\bf 48}, 785 (1993).
\bibitem{Kim3} See Appendix A of Ref.~\cite{Kim2}.
\bibitem{Mao} J.-m.\ Mao and B.\ Hu, J.\ Stat.\ Phys.\ {\bf 46}, 111 (1987)
\end{references}
\end{document}